\documentclass[12pt]{article}
\usepackage[active]{Srcltx} 
\usepackage{graphics}
\begin{document}

\begin{titlepage}

\begin{flushright}
Bicocca-FT-99-39\\ MIT-CTP-2924\\ NSF-ITP-99-136\\[1mm] {\tt
hep-th/9911113}
\end{flushright}

\begin{center}
\title{Monopoles in String Theory}

\author{Amihay Hanany}
\author{Alberto Zaffaroni}
\hfill \vskip .4in {\large\bf Monopoles in String Theory}
\end{center}
\vskip .4in
\begin{center}
{\large Amihay Hanany$^{a,b}$ and Alberto
Zaffaroni$^{c,}$\footnotemark} \footnotetext{e-mail:
hanany@mit.edu, alberto.zaffaroni@mi.infn.it} \vskip .1in (a){\em
Institute for Theoretical Physics, Santa Barbara, CA 93106, USA.}
\vskip .1in (b){\em Massachusetts Institute of Technology,
Cambridge MA 02139, USA\footnotemark} \footnotetext{Permanent
Address} \vskip .1in (c){\em I.N.F.N. - Sezione di Milano, 20133,
Italy.}
\end{center}
\vskip .4in
\begin{center} {\bf ABSTRACT} \end{center}
\begin{quotation}
\noindent
A realization of $E_{n+1}$ monopoles in
string theory is given. The NS five brane stuck to an Orientifold
eight plane is identified as the 't Hooft Polyakov monopole.
Correspondingly, the moduli space of many such NS branes is
identified with the moduli space of $SU(2)$ monopoles. These
monopoles transform in the spinor representation of an $SO(2n)$
gauge group when $n$ D8 branes are stacked upon the orientifold
plane. This leads to a realization of $E_{n+1}$ monopole moduli
spaces. Charge conservation leads to a dynamical effect
which does not allow the NS branes to leave the orientifold plane.
This suggests that the monopole moduli space is smooth for $n<8$.
Odd $n>8$ obeys a similar condition. Using a chain of dualities,
we also connect our system to an Heterotic background with
Kaluza-Klein monopoles.
\end{quotation}
\vfill November 1999
\end{titlepage}
\eject
\noindent

\newcommand{\Tr}{\mbox{Tr\,}}
\newcommand{\Dirac}{/\!\!\!\!D}
\newcommand{\beq}{\begin{equation}}
\newcommand{\eeq}[1]{\label{#1}\end{equation}}
\newcommand{\bea}{\begin{eqnarray}}
\newcommand{\eea}[1]{\label{#1}\end{eqnarray}}
\renewcommand{\Re}{\mbox{Re}\,}
\renewcommand{\Im}{\mbox{Im}\,}

\section{Introduction}
String theory has many realizations of the monopole solutions of
't Hooft and Polyakov. Basically, whenever there is an Abelian
gauge field which has the potential of being a subgroup of a
non-Abelian gauge group then we would expect an object which will
be the analog of the monopole solution in field theory.

D branes offer a convenient framework for realizing monopole
configurations in supersymmetric gauge theories and studying their
moduli space. In a standard construction
\cite{Strominger,Diaconescu,HW}, monopoles for a non-Abelian gauge
theory living on the world-volume of a set of branes are
identified with other branes, of different type, ending on them.
The type and the extension in space-time of the branes are chosen
in order to have a BPS configuration. An overview of the various
possibilities is described in Section 2. $SU(N)$ monopoles of
magnetic charge $k$ are easily described in this context and
orthogonal and symplectic groups can be studied by introducing
orientifold planes. Much more difficult is the study of $E_n$
groups. In this paper we provide an explicit example of brane
configurations that are naturally interpreted as $E_n$ BPS
monopoles.

$E_n$ gauge groups, where $n$ runs from 1 to 8, can be described
in the Type I$^\prime$ string theory using backgrounds where the
dilaton is blowing up at some orientifold plane. These backgrounds
with $E_n$ symmetry are in one-to-one correspondence with the
point in moduli space of the nine-dimensional Heterotic theory
where there is a perturbative enhanced symmetry. The spectrum of
electrically charged states in these  Type I$^\prime$ models has
been extensively discussed in the literature, since it provides a
non trivial check of string dualities. In particular, D0 branes,
stuck on the orientifold plane,
 have been identified with
the electrically charged states that become massless at the point
in moduli space where an $SO(2n)\times U(1)$ gauge symmetry
realized on D8 branes and Abelian bulk fields is enhanced to
$E_{n+1}$. Much less attention has been paid to magnetically
charged objects. We will show that stuck NS branes  are naturally
interpreted as monopoles. Together with systems of D6 branes
stretched between the D8 branes responsible for the
 $SO(2n)$ symmetry, these NS branes
give a stringy description of $E_{n+1}$ monopoles.

In the absence of D8 branes, we obtain a description of $k$
$SU(2)$ monopoles in terms of $k$ NS branes moving on the
orientifold plane. The moduli space of such monopoles is known to
be smooth. Our NS branes are, by construction, stuck on the
orientifold plane. For living outside the plane, a NS brane needs
an image under the orientifold projection and, in principle, two NS
branes could meet and move outside in the bulk. An obvious
singularity would show up in the moduli space of the NS brane if
they could leave the orientifold plane. We will see that a charge
conservation argument does not allow them to leave. This is the
string theory explanation for the smoothness of the monopole
moduli space. The very same construction, generalized to the
presence of D8 branes, predicts that the $E_{n+1}$ monopole moduli
space is also smooth.

The existence of monopoles in string theory can be also addressed
from a different perspective. They can show up quite naturally in
string compactifications. Sen realized that KK monopoles in the
Heterotic  string can be identified with BPS monopoles of a
spontaneously broken gauge symmetry \cite{Sen}. The same system
was recently studied in order to get the exact moduli space of the
Heterotic string near an ALE singularity without gauge field
\cite{Witten}. We will see that these Heterotic configurations are
connected with the ones discussed in this paper by a chain of
dualities. This connection is a further evidence of our
identification of stuck NS branes with monopoles. On the other
side, our construction can be used to explain and generalize the
result in \cite{Sen,Witten}. Various papers have appeared which
tried to connect the Heterotic moduli space with the Coulomb
branch of three-dimensional N=4 gauge theories
\cite{rozali,plesser,mayr}. This connection is expected due to the
relation between the Coulomb branch of some three-dimensional N=4
gauge theories and monopole moduli spaces \cite{HW}. Our point of
view in this paper is that we look for the existence and the
identification of the relevant monopoles in the problem, leaving
three dimensional gauge theories aside.

This paper is organized as follows. In Section 2, we give an
overview of the various string theory realizations of monopoles.
Section 3 contains the description of the Type I$^\prime$
configuration that is the object of this paper and the explicit
realization of the $E_{n+1}$ monopoles. Section 4 makes a
connection with the KK monopoles in the Heterotic string. Various
comments on the smoothness of moduli space are contained in
Section 5. Section 6 contains a brief discussion of the $D_k$
case.

\section{Monopoles in String Theory}
\label{mon}
Examples of monopoles in string theory have
been studied in great detail and here we summarize some of the
cases\footnote{For older realizations of monopoles in string theory
see \cite{kiritsis}.}.

Probably the most studied example is the one realized by a D1
brane which is stretched between a pair of two D3 branes
\cite{Strominger,DouglasLi,Green, Tseytlin,Diaconescu}. This is
the classical example for a spontaneously broken four dimensional
N=4 supersymmetric YM theory with gauge group SU(2), the one which
was explicitly used in the solutions of 't Hooft and Polyakov.
There are several natural realizations of such monopole solutions.
A Dp brane stretched between a pair of Dp+2 branes will be a $p-1$
brane solution in a $p+3$ dimensional theory, for $p\leq6$. The
classical solution of the Higgs field represents the shape of this
brane configuration \cite{DouglasLi}. This family of solutions is
related to the case $p=1$ by a set of T-dualities. The field
theory interpretation of such dualities is dimensional reduction
starting from a higher dimensional theory and reducing some of its
dimensions. The case $p=0$ is somewhat special and represents an
object which has the interpretation of an instanton in 3
dimensions. It is identified with the Euclidean monopole solution
of Polyakov \cite{Polyakov} and is represented by a Euclidean D0
brane.

Another example of monopoles in string theory is realized by
applying S-duality to the case $p=3$. This gives a configuration
of a D3 brane stretched between a pair of NS five branes, a
configuration which was well studied \cite{HW}.  One can apply a
further SL(2,Z) transformation on this configuration and get a D3
brane stretched between a pair of (p,q) five branes. \cite{HW}
also demonstrated a connection to three dimensional gauge theories
with N=4 supersymmetry by mapping the Coulomb branch moduli space
to certain monopole moduli spaces, a problem which was motivated
by field theory studies, for the $SU(2)$ case in \cite{SW3} and
for $SU(n)$ in \cite{CH}.

One other example is that of \cite{Sen} in the study of the
Heterotic string on a Taub-NUT space. Sen finds that the monopole
solution in the form of a Kaluza Klein monopole is interpreted as
a 't Hooft Polyakov monopole. The radius of the circle in the
Taub-NUT space plays the role of the scalar VEV in the
spontaneously broken $SU(2)$ group which becomes enhanced when the
radius approaches the self dual radius.

A common feature to all of these string theory configurations is
the fact that they all represent classical monopole solutions.
Correspondingly the moduli space of solutions for these objects
coincides with the moduli space of monopoles \cite{AH}. The
structure of the moduli space solution for the simplest case of 2
monopoles in an $SU(2)$ gauge theory was studied in detail in the
book of Atiyah and Hitchin. The space is known as the Atiyah
Hitchin space. It has dimension 4 and admits a hyperKähler metric.
We intuitively identify the 4 parameters as the 3 relative
positions of the monopoles and an angle associated with their
relative phase. The Atiyah Hitchin metric consists of the
following three types of contributions. There is a natural
expansion parameter for the monopole moduli space. It is given by
the distance between the two monopoles as measured in units of
inverse scalar VEV. A typical metric component has an expansion of
the form \cite{mon}
\begin{equation} 1-{{\rm const}\over \langle\phi \rangle x} + O(e^{-\langle\phi\rangle x}) \label{e1}\end{equation}
where $x$ is the monopole separation and $\langle\phi\rangle$ is
the Higgs VEV.
For very large separations the moduli space
looks flat with two orbifold singularities. There is a correction
to this metric coming with an inverse power of the separation
between the monopoles. This changes the space to a circle bundle
over $R^3$. There are further exponential corrections to the
metric which make this space smooth.

It is interesting to consider how such exponential corrections
arise in each of the examples discussed above. We will mention the
corrections for the moduli space of two $SU(2)$ monopoles in each
of these cases. For a pair of Dp branes stretched between a pair
of Dp+2 branes one can stretch a fundamental string which is
bounded by both the Dp and the Dp+2 branes. This gives rise to a
worldsheet instanton with an action $x\langle\phi\rangle$ where
$x$ is the separation between the two Dp branes interpreted as the
separation between the two monopoles and $\langle\phi\rangle$ is
the scalar VEV of the spontaneously broken $SU(2)$, measured as
the distance between the two Dp+2 branes in units of $l_s^2$.

The case of a pair of  D3 branes stretched between two NS branes
gives rise to a Euclidean D1 brane bounded between both the D3
branes and the NS branes. Here $\langle\phi\rangle$ is the
distance between the two NS branes measured in units of
$g_sl_s^2$. Similarly for a pair of D3 branes stretched between a
pair of $(p,q)$ five branes the exponential correction is due to a
$(p,q)$ string bounded by both the three branes and the five
branes. $\langle\phi\rangle$ is the distance between the two
$(p,q)$ five branes measured in units of the $(p,q)$ string
tension.

The purpose of this paper is to discuss a certain class of Type
I$'$ theories. In the spirit of the introduction above, whenever
we observe a gauge group we would like to look for the monopole
solutions for this theory. One guide line for this search will be
what we call a ``generalized Montonen Olive duality." This duality
in one of its forms states that the monopole spectrum of a four
dimensional N=4 supersymmetric gauge theory with gauge group $G$
sits on the lattice of the electric spectrum of another four
dimensional N=4 supersymmetric gauge theory with gauge group
$\widetilde{G}$. $\widetilde{G}$ is defined as having a root
lattice which is the dual of the root lattice of $G$. The extension of
the Montonen Olive duality that we will use is in dimensions different
than 4. We will say that the monopole spectrum of a gauge theory
with 16 supercharges in $p+3$ dimensions and a gauge group $G$
sits on the same lattice as the electric spectrum for a gauge
theory with 16 supercharges in $p+3$ dimensions and gauge group
$\widetilde{G}$. This property will be used when discussing some
particular backgrounds with various groups $G$.

\section{Monopoles in Type I$'$ String Theory}

Type I$'$ in its most simple form is described by a background
with $R^9\times S^1/Z_2$ where the $Z_2$ acts on the circle by
reflection of the coordinate while changing the orientation of the
string worldsheet. It is convenient to think of the circle as an
interval. There are two fixed planes under the $Z_2$ action which
carry a D8 RR charge of magnitude -8. These planes are called
orientifold planes and are denoted by $O8^-$. Charge conservation
on the interval implies a constraint on the number of physical D8
branes, which carry RR charge 1, to be 16. There is an option to
replace one of the orientifolds by a positively charged
orientifold, denoted by $O8^+$, with RR charge +8. In this case
charge conservation on the interval implies that there are no
physical branes in the bulk of the interval. The second option,
having no physical D8 branes, will be less interesting for the
applications in this paper and will not be considered.

The Type I$'$ string coupling, or more precisely its inverse, is a
varying function on the positions of the various D8 branes. It
satisfies a Laplace equation in 1 space dimension with delta
function sources at the positions of the D8 branes on the
interval, together with the negative charge source at the
boundaries of the interval. As such it is a piece-wise smooth
linear function. Denote the positions of the D8 branes on the
interval by $x_i$ and the coordinate on the interval by $x$, then
the string coupling obeys the following equation
\begin{equation}
\frac{l_s}{g_s(x)}=\sum_{i=1}^{16}|x-x_i|-8|x|-8|R-x|+\frac{l_s}{g_s^0}.
\end{equation}
When this function vanishes the string coupling diverges and some
states in the string spectrum become massless. Such a vanishing
can happen only at the boundaries of the interval since it is a
positive function. As discussed in \cite{BGL} a half
\footnote{Half a brane in this case means that the brane carries
half of the charge of a physical brane. Away from the orientifold
a physical brane consists of a half brane and its image under
space reflection. On the orientifold, half a brane can exist with
no images.} D0 brane which is stuck to one of the orientifold
points becomes massless as the string coupling diverges. This
state together with a massless half anti D0 brane and the gauge
field which sits in the string coupling multiplet form a gauge
group $SU(2)$. One can describe this process as an inverse Higgs
mechanism for getting an enhanced $SU(2)$ gauge group. The scalar
VEV is given by the D0 brane mass,
\begin{equation}
\langle\phi\rangle={1\over 2g_s l_s}.
 \end{equation}
 The gauge
coupling of this gauge theory is given by the usual coupling on a
D8 brane, \begin{equation}{1\over g_{YM}^2}={1\over g_s
l_s^5}.\end{equation}

At this point we would like to ask what is the object which can be
identified with the 't Hooft Polyakov monopole for this
spontaneously broken $SU(2)$. Since it is a BPS state it is enough
to know its mass in order to identify its charge. On the other
hand we can apply the classical formula for the tension of the
monopole, $T_{mon}$ from its field theory value,
\begin{equation}
T_{mon}=\frac{\langle\phi\rangle}{g_{YM}^2}={1\over
2g_s^2l_s^6}.
\end{equation}
This formula identifies the monopole as a half NS brane which is
stuck to the $O8^-$ plane.\footnote{One may question the existence
of half NS brane on an $O8^-$ plane. We thank Oren Bergman for
raising this issue. However, its existence is imposed by the
``generalized Montonen Olive duality" principle.} More details on
this identification appear in section \ref{het}.

The system of NS branes and D8 branes is the same system which was
studied in the context of six dimensional gauge theories in
\cite{BK} and \cite{HZ4}. For definiteness we will take the NS
branes to lie along 012345 directions, the D8 to be point like in
direction 6 and, when present, D6 branes will span the coordinates
0123456.

This picture can be generalized to include additional physical D8
branes sitting at the $O8^-$ plane. For finite string coupling the
world volume theory of $n$ D8 branes stacked at the orientifold
plane is $SO(2n)$. Together with the $U(1)$ gauge field, which is
in the multiplet which contains the dilaton, this symmetry is
enhanced to $E_{n+1}$ for a diverging dilaton at the fixed point
of the interval. We may ask where are the extra states in the
adjoint representation of the $E_{n+1}$ theory. This question was
analyzed by \cite{BGL} who argued that a half D0 brane stuck at
the $O8^-$ plane transforms under the spinor representation of
$SO(2n)$ and is naturally charged with respect to the $U(1)$ with
a charge $\frac{1}{2}.$ It is amusing to note that such
identification requires a non-trivial bound state of two half D0
branes for the cases $E_7$ and $E_8$ \cite{BGL}. To find
this bound state is a
non-trivial problem in quantum mechanics of such stuck D0 branes.

With this picture in mind we now turn to the spectrum of monopoles
in such an $E_{n+1}$ theory. At this point we recall the
``generalized Montonen Olive duality" principle which was
mentioned at the previous section. As the $E_{n+1}$ root lattice
is self dual, the monopole spectrum sits in the adjoint
representation of the $E_{n+1}$ group. This decomposes naturally
to monopoles in the adjoint representation of the $SO(2n)$ group
and neutral under the $U(1)$ and to monopoles in the spinor
representation of $SO(2n)$ charged under the $U(1)$. The BPS
objects in the adjoint representation of $SO(2n)$ are monopoles
for the D8 branes and are naturally given by $D6$ branes stretched
between a pair of neighboring D8 branes. This is one of the cases
which was mentioned in the previous section for $p=6$. What about
the BPS objects in the spinor representation? The $U(1)$ charge of
the object and its mass formula lead to identify it with a half NS
five brane stuck at the $O8^-$ plane. We may ask why it transforms
in the spinor representation of $SO(2n)$. This is not a simple
problem and we may only give some suggestions. According to \cite{HZ4}
there is a linking number \cite{HW} which is induced by the D8
brane on the stuck NS brane. This linking number is half the unit
of charge for a D6 brane. One may ask if there is a state
associated with this charge. It must be a supersymmetric singlet
as there are no additional massless multiplets which are induced
by the D8 brane on the NS brane. It is highly suggestive that
quantization of such objects leads the NS brane to transform under
the spinor representation of $SO(2n)$. It is not easy to show
this, though. It is again amusing to note that for the cases of
$E_7$ and $E_8$ groups the analysis above suggests that there will
be non-trivial bound states of a pair of half NS branes which are
required to complete the adjoint representation of the gauge
group. It is an interesting problem to try and show how this
arises.

\subsection{Smoothness Puzzle}
\label{smooth} Let us look more carefully at the case of a gauge
group $E_1$. This corresponds to an $O8^-$ plane with a diverging
string coupling at the enhanced point and a finite string coupling
for the spontaneously broken phase. Let us take two half NS branes
stuck on the $O8^-$ plane. According to our identification these
two branes are two monopoles in an $SU(2)$ gauge group.
Correspondingly their moduli space, which consists of 3 relative
positions inside the $O8^-$ plane and a relative phase angle in
the eleventh direction, is identified as the (reduced) moduli
space of 2 $SU(2)$ monopoles or the Atiyah Hitchin manifold. The
manifold is known to be smooth. This points to some difficulty.
There is apparently a singularity in the moduli space of two half
NS branes as they move inside an $O8^-$ plane. They can approach
each other along the three directions inside the $O8^-$ plane and
leave to the bulk of the interval as a pair of a brane and its
image under the reflection of the Type I$'$ interval. The motion
away from the orientifold plane corresponds to a scalar in a
massless tensor multiplet. This tensor multiplet does not exist as
a massless state in the phase in which the NS branes are stuck to
the orientifold. Correspondingly, there is a singular point in the
hypermultiplet moduli space in which the tensor multiplet becomes
massless. This is in contradiction with the smoothness of the
moduli space of two monopoles!

\begin{figure}[h]
  \centering
  \resizebox{12cm}{!}{\includegraphics{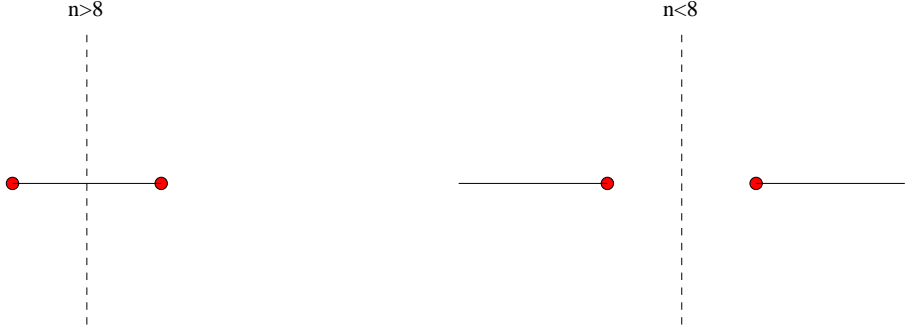}}
  \caption{A physical NS brane in the presence of an $O8^-$ plane.
  The dashed line represents an $O8^-$ plane with a stack of $n$ D8
  branes and the circle denotes
  half a physical NS brane. The solid lines are D6 branes. There are two cases to consider
  depending on the number $n$. For $n>8$ the D6 branes stretch to the
  $O8^-$ plane while for $n<8$ they stretch away from it. }\label{tails}
\end{figure}

How can this be solved? We recall that the Type I$'$ has
irregularities in the form of a cosmological constant in an
intermediate region between two D8 branes or between D8 brane and
an $O8^-$ plane.
We use units where a D8 brane induces a jump in the cosmological constant
of magnitude 1.
For our example of the $E_1$ theory the cosmological constant outside
the $O8^-$ plane is -8. As discussed in \cite{HZ4}, a NS brane in a non-zero
cosmological constant background has D6 brane tails.
Charge conservation for the Ramond-Ramond fields in the presence
of a cosmological constant of magnitude $m$ implies a relation between
the numbers $n_R,n_L$ of D6 branes that end on a NS
brane from the right and from the left~\cite{HZ4},
\begin{equation}
n_L-n_R=m
\label{cos}
\end{equation}
In this formula, left and right refer to the picture where the Type I$'$
background is represented as a segment limited by two $O8^-$ planes.
As discussed in \cite{HZ4}, whenever equation~(\ref{cos}) is satisfied
the six-dimensional theory living on the D6 and NS brane system is
anomaly free.

For the case of a NS brane near an $O8^-$ plane, the cosmological
constant is -8 and, according to equation~(\ref{cos}), we have 8 half
D6 branes stretching away from the $O8^-$ plane, as in figure
\ref{tails} for the case $n=0$. It is crucial that the D6 branes
are stretched away from the $O8^-$ plane. This is the key point
which resolves our puzzle. Suppose that a pair of stuck half NS
branes meet and attempt to move away from the $O8^-$ plane. This
is not possible due to energetic reasons. Long D6 branes can not
be formed as soon as the NS branes leave to the bulk of the
interval. We conclude that the pair of half NS branes can not
leave the $O8^-$ plane. Instead they are confined to the plane and
there are no singular points associated to the motion outside. The
moduli space in question is smoothed out. In some sense this
effect can be thought of as a higher order effect. The classical
moduli space of two half NS branes has a flat metric with an
orbifold singularity,
\begin{equation}
\frac{ R^3\times S^1 }{Z_2}.
\label{classical}
\end{equation}
The singularity at the origin is interpreted as the point at which
naively a pair of half NS branes meet and attempt to leave the
$O8^-$ plane. This singularity is smoothed out by an exponential
correction to the metric coming from Euclidean D0 branes, which
are stretched between the two half NS branes, of action
$\frac{x}{g_sl_s}.$ Here $x$ is the distance between the NS
branes, $g_s$ is the string coupling at the $O8^-$ plane and $l_s$
is the string scale. These objects prevent the NS branes from
meeting on the $O8^-$ plane. We can restate this in terms of the
mass for a tensor multiplet. Classically, there is a massless
tensor multiplet which arises as the NS branes meet and leave the
$O8^-$ plane. This tensor multiplet gets exponential contributions
to its mass from instantons coming from Euclidean D0 branes
stretched between the two half NS branes.

It is easy to generalize this picture to a higher number, $n$, of
D8 branes sitting on the $O8^-$ plane. As long as $n<8$ in figure
\ref{tails} the charge of the combined system, $O8^-$ and $n$ D8
branes, is negative. Consequently, half physical NS branes are
confined to this system and will not leave to the bulk of the
interval as a pair.

The following statement can be used as a prediction on the
behaviour of the moduli space of monopoles for several gauge
groups. Let us first consider the case $E_1$, namely an $O8^-$
plane. $k$ half NS branes stuck on this plane correspond to $k$
$SU(2)$ monopoles. Their moduli space is identified with the
moduli space of $k$ $SU(2)$ monopoles. It is a $4k$ dimensional
space which is believed to be smooth. (There is a 4 dimensional
trivial part which is associated to the center of mass motion for
the monopoles but the remaining space is believed to be smooth).
The case $k=2$ was calculated explicitly by Atiyah and Hitchin and
was shown to be smooth. Our resolution to the puzzle actually
supports the claim that the $k$ $SU(2)$ monopole moduli space is smooth
for any $k$.

We can further extend this claim for $E_{n+1}$ models. Here the
statement will only deal with a particular part in monopole moduli
space of $E_{n+1}$ which consists only of those monopoles in the
spinor representation of $SO(2n)$ under the decomposition
$SO(2n)\times U(1)\subset E_{n+1}$. The moduli space of $k$ such
monopoles is smooth for any $k$. There are no singularities
associated with two such monopoles meeting and leaving to the bulk
of the interval.

\subsection{$n>8$ and an Odd Puzzle}

\begin{figure}[h]
  \centering
  \resizebox{8cm}{!}{\includegraphics{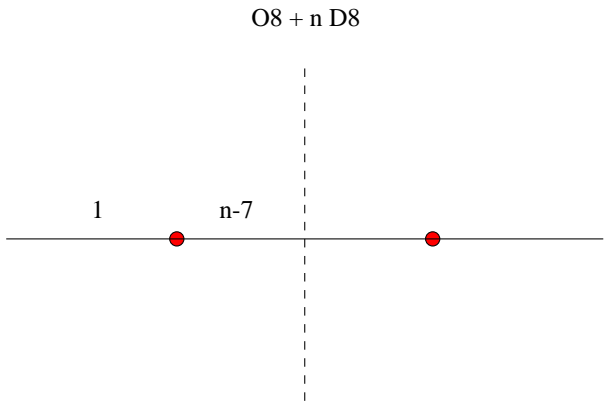}}
  \caption{A physical NS brane in the presence of an $O8^-$ plane
  and an odd number, $n$ of D8 branes. There are $n-7$ half D6 branes
  stretching between the half NS brane and its image. There is a half
  D6 brane stretching away from the half NS brane.}\label{odd}
\end{figure}

What about the case of $n$ D8 branes that are enough to make the
charge of the $O8^-$ plane positive? In this case a pair of half
NS branes is allowed to leave the plane and a total of $n-8$ half
D6 branes can stretch in between the two half NS branes. Here we
face another puzzle which first appears for the case $n=9$. As
already pointed out in \cite{hzsix}, consistency with tadpole
cancellation restricts the number of D6 branes which cross the
$O8^-$ plane to be even. This is not the case for $n-8$ odd. One
would conclude that $n>8$ odd is not a consistent brane
configuration. However, there is another alternative for a
physical NS brane in the bulk which is consistent with this
restriction. This is the alternative which was discussed in
\cite{hzsix}. For $n$ odd, one can stretch, as in figure
\ref{odd}, $n-7$ half D6 branes in between the half NS brane and
its image and a single half D6 brane away from the half NS brane.
This configuration is consistent with charge conservation for the
RR 6 brane charge in the presence of a non-zero cosmological
constant.

A brane which is stretched away from the NS brane poses the same
problem that we had for the $n<8$ case. Suppose that in the
background of $n$ odd D8 branes stacked upon an $O8^-$ plane there
is a pair of stuck half NS branes. Such a pair can not leave the
$O8^-$ plane to the bulk of the interval. It is not allowed by the
same energetic reasoning as in the case for $n<8$. For even $n>8$
there is no need to stretch a D6 brane away from the NS brane and
therefore a pair of half NS branes is allowed to leave the $O8^-$
plane. It is not clear how to interpret this phenomenon from the
point of view of monopoles in the corresponding gauge theory.

\subsection{A massless Three Brane?}
\label{threebrane}

Singularities in moduli space are typically associated with the
appearance of some massless states, or more generally tensionless
objects. The classical moduli space of two half NS branes,
(\ref{classical}), is singular. Here classical means with respect
to the natural expansion parameter which is determined from the
form of the Atiyah Hitchin manifold. In our case it is the
Euclidean D0 brane action $\frac{x}{g_sl_s}$, with $x$ the
distance between the two NS branes. We may ask what is the object
which becomes massless/tensionless in the limit of a small string
coupling. It is easy to find that this object is given by a D4
brane which is stretched between the two NS branes \footnote{One
may actually need to consider a pair of half D4 branes in order to
avoid problems of existence for such a configuration. The
discussion is not affected by this, though.}. This is the only
object which can stretch between two NS branes inside an $O8^-$
plane in a supersymmetric fashion. For small expansion parameter
the D4 brane gives rise to a tensionless three brane as the pair
of half NS branes coincide. This is clearly a naive picture since
the moduli space gets smooth by exponential corrections in the
expansion parameter. Correspondingly the three brane gets
exponentially small corrections to its tension of the order of
$\exp(-\frac{x}{g_sl_s})$. This three brane actually never gets
tensionless as the moduli space is smooth. It is amusing to note
that this object maps in the Heterotic string language, which will
be discussed in the next sections in detail, to a small instanton
which wraps a vanishing two-cycle.

\section{Heterotic String Moduli Space}
\label{het}
In this Section, we compare the identification of
$E_{n+1}$ monopoles with configurations of D6/D8/NS branes by
performing a series of T/S dualities. At the same time, we make
contact with different approaches, where monopole moduli spaces
appear in the perturbative description of the Heterotic string.

Starting with the configuration discussed in the previous Section,
a T duality along the 6 direction brings us to the Type I theory.
At this point an S duality leads to the Heterotic string, where
the enhanced gauge symmetry phenomenon of the $E_{n+1}$ theory can
be perturbatively studied. In this process, the D8 branes are
mapped to D9 branes in the Type I theory and the $SO(32)$ gauge
fields in the Heterotic string, while the $k$ NS branes are mapped
into $k$ Kaluza-Klein monopoles (this is the same as a Taub-NUT
space) of the Type I or Heterotic theory. If the identification of
the Type I$^\prime$ stuck NS branes as monopoles is correct, we
expect that the moduli space of monopoles appears as the space of
vacua of the Heterotic string defined on a Taub-NUT space. It was
shown in \cite{Sen,Witten} that the moduli space of $k$ $SU(2)$
monopoles indeed appears in this context. More precisely, Sen
argued that the Kaluza Klein monopole is identified with a 't
Hooft Polyakov monopole of the $SU(2)$ gauge group which is
enhanced at the self dual radius. Studying the problem in the limit where the
Taub-NUT space reduces to an ALE space, Witten conjectured a relation to
three dimensional gauge theories. The connection between the two
points of view
is made by realizing that the monopole moduli space coincides with the
Coulomb branch of the three dimensional gauge theory \cite{CH,HW}.

We can formulate the problem, from the point of view of the
Heterotic string, as follows. We are interested in the moduli
space of hypermultiplets in the Heterotic string on $K_3$. In
order to simplify the problem and keep only few hypermultiplets,
we replace $K_3$ with a non-compact space obtained by zooming on
local singularities of the manifold. This is equivalent to
considering the Heterotic string defined on an ALE space. This
configuration can be continuously deformed to the one we obtained
after T and S duality; a Taub-NUT space looks near infinity like
a non-trivial $S^1$ bundle over $R^3$
 and reduces indeed to an ALE space in the limit where the
radius of the $S^1$ becomes large. As a result of the zoom, the 6
dimensional effective theory describing the moduli for this
``compactification'' is decoupled from gravity and bulk modes.
 We want to focus on the moduli space
corresponding to the 6 dimensional modes. Other parameters, which
specify the background, such as the radius of the Taub-NUT space
or Wilson lines for the 10 dimensional gauge fields, are not
dynamical, since they are VEVs of 10 dimensional decoupled fields.
Due to the decoupling of gravity, the 6 dimensional hypermultiplet
moduli space is a hyper-K\"ahler manifold. The hypermultiplet
moduli space is not corrected by string loops, but
 receives $\alpha^\prime$ corrections.
The classical moduli
space is generically singular. In backgrounds with non-trivial
gauge bundles, some singularities, for example those associated
with small instantons, survive quantum corrections and find their
explanation in non-perturbative phenomena. In the trivial bundle
case, there are examples where quantum corrections
 smooth the classical singularity
\cite{Sen,Witten}. In these examples, quantum corrections
reproduce the expected moduli space of monopoles.

\subsection{The moduli space of KK monopoles}
Let us review what is known about the moduli space of KK monopoles
in the Heterotic string with trivial gauge fields
\cite{Sen,Witten}. These results strongly support our
identification of stuck NS branes in the dual Type I$^\prime$
theory as BPS monopoles.

Consider
 the dynamics of $k$ Kaluza-Klein monopoles
 in the Heterotic theory,
or, in other words, a multi-centered Taub-NUT space. Their exact
moduli space has been found by Sen \cite{Sen}.

The Taub-NUT space has a metric
\begin{equation}
ds^2= V(\vec x)d\vec x^2+V^{-1}(\vec x)(dx_4+\vec\omega\times d\vec x)^2
\label{pot}
\end{equation}
which is completely specified by the potential
\begin{equation}
V(\vec x)=1+\sum_{i=1}^k{1\over |\vec x-\vec x_i|} \label{pottwo}
\end{equation}
In the perturbative Heterotic string, the moduli space
corresponding to this ``compactification'' is parameterized by the
expectation values of $k-1$ six dimensional hypermultiplets. The
scalar components of these hypermultiplets are obtained by
reducing the metric and the B field along the $k-1$ two-cycles of
the space. In the dual Type I$^\prime$ description, these
hypermultiplets live on the world-volume of the stuck NS branes
and parameterize their position in the space transverse to it. The
classical moduli space is a symmetric product of $k-1$ copies of
$R^3\times S^1$ and  is singular. The singularity is the effect of
considering the low energy supergravity; it is smoothed out by
higher derivative corrections. This can be understood as follows
\cite{Sen}. By varying the radius R of the $S^1$, the $SO(32)$
Heterotic string can be driven to the self-dual point
($R^2=\alpha^\prime$) where there is a perturbative enhanced
$SU(2)$ symmetry. The $W^\pm$ bosons responsible for enhancing the
symmetry from $U(1)$ to $SU(2)$  are perturbative BPS string
states with $(n,m)= \pm (1,-1)$ units of momentum and winding
along $S^1$, as can be seen from the BPS formula
\begin{equation}
M^2_{BPS}=({n\over R}+{mR\over\alpha^\prime})^2. \label{ee1}
\end{equation}
The Kaluza-Klein monopoles are identified as BPS monopoles of the
$SU(2)$ enhanced symmetry. A KK monopole has indeed the same
magnetic charge $(1,-1)$ with respect to the $S^1$ momentum and
winding of a BPS monopole. That a KK monopole carries one unit of
momentum is part of the definition of the object. That it also
carries $-1$ unit of winding follows from the equation
\begin{equation} dH= \Tr F\wedge F-\Tr R\wedge R \label{ee2}\end{equation} that
needs to be satisfied in any consistent Heterotic model. It
follows from this identification that the exact moduli space for
$k$ KK monopoles is the moduli space of $k$ BPS monopoles of
$SU(2)$. This is expected to be a smooth manifold; for $k=2$ it is
the Atiyah-Hitchin manifold.

Monopoles exist in the phase where the group is spontaneously
broken to $U(1)$ by the expectation value of some Higgs field. The
metric on the moduli space has an expansion in terms of the
product $\langle\phi\rangle x$ of the monopole distance with the
Higgs field VEV which is given in formula~(\ref{e1}) of Section
~\ref{mon}.
 We can identify the gauge theory
parameters with the Heterotic ones as follows. The 9 dimensional
$SU(2)$ group has coupling constant $Re^{-2\phi_h}$, where
$\phi_h$ is the Heterotic Dilaton. It follows from formula
(\ref{ee1}) that the VEV of the field responsible for the
spontaneous symmetry breaking is related to the $S^1$ radius by
$R/\alpha^\prime -1/R$. We see that the expansion of metric
components in equation~(\ref{e1}) can be interpreted, in the large
radius limit, as the $\alpha^\prime/R^2$ expansion.  For large
$R^2/\alpha^\prime$ we recover the supergravity result, a flat
singular metric. The perturbative expansion only contains a
one-loop (in $\alpha^\prime$)
 contribution, which comes
with a negative sign. Only after including the non-perturbative
corrections due to instantons the metric becomes
smooth.

A second example, due to Witten \cite{Witten}, deals with the
moduli space of the Heterotic string near an ALE singularity with
no gauge bundle. For a $Z_k$ singularity, the relevant 6
dimensional fields are $k-1$ hypermultiplets parameterizing the
blowing-up modes of the manifold. Combining symmetries,
semi-classical arguments and the assumption of smoothness, the
moduli space for $k=2$ was unambiguously identified in
\cite{Witten} with the Atiyah-Hitchin manifold. It was then
conjectured that the moduli space for a singularity of type $G$ is
the Coulomb branch of a three dimensional gauge theory with 8
supercharges and gauge group $G$. Due to the relation between
monopole moduli spaces and three dimensional gauge theories
\cite{CH}, this result agrees with Sen's one. This example can be
indeed considered as a limit of the previous one, where the radius
$R$ is sent to infinity while keeping the blowing-up parameters
fixed; this scaling preserves the form of the metric. Notice that
the $SU(2)$ group is somehow hidden in this approach.

The expansion of the metric components in equation~(\ref{e1}) in
this example is just the $\alpha^\prime$ expansion. There is a
world-sheet one-loop correction to the classical singular metric
and a series of world-sheet instanton corrections. Notice that a
world-sheet instanton in this case is a fundamental string wrapped
over a two-cycle of the ALE space.
 An S-duality transforms it into a D1 brane
of the Type I theory and a T-duality into an Euclidean D0 brane
stretched between the NS branes in the Type I$^\prime$
description. This is in agreement with the identification made in
Section \ref{smooth} of D0 branes as responsible for the
corrections to the classical metric. As discussed in section
\ref{threebrane}, we can here mention again the existence of a BPS
three brane given by a small Heterotic instanton which wraps a
two-cycle of the ALE space. Classically, when the metric is
singular, it is tensionless. As one includes $\alpha'$ corrections
the tension gets exponential corrections and does not vanish
anywhere on the moduli space.

The question of smoothness of the moduli space for the Heterotic
on ALE is not as clean as in the Type I$'$ picture. It is not
clear how to translate the statement on charge conservation made
in section \ref{smooth}. It is also not clear how to translate the
statement that the NS branes are confined to the $O8^-$ plane to
the Heterotic picture. Such questions, if answered, would give us
a better perspective on the issue of smoothness of the moduli
space from the Heterotic string point of view.

\subsection{Generalization to the $E_{n+1}$ case}
The generalization of Sen's example to the $E_{n+1}$ case is simple. In the
Heterotic language, we can tune the radius and the Wilson lines in
such a way that an $E_{n+1}$ symmetry in enhanced. The BPS
formula, for generic Wilson lines, reads
\begin{equation} M_{BPS}^2=\left
({n-A\cdot P-mA^2/2\over R}+{mR\over \alpha^\prime}\right )^2
\eeq{ee4} where $A$ is a sixteen component vector representing the
Wilson line and $P$ is an element of the $SO(32)/Z_2$ lattice. In
Type I$^\prime$ picture, an entry $0$ in the 16 component vector
$A$ corresponds to a D8 brane sitting at the orientifold plane
where the coupling constant is blowing up, while an entry $1/2$
corresponds to a D8 brane at the other orientifold. $E_{n+1}$ is
obtained by choosing a vector of the form
$A=(0,0,....,0,1/2,...1/2)$, with $n$ entries equal to zero. The
critical radius is
$R^2=\alpha^\prime(1-A^2/2)=\sqrt{\alpha^\prime(8-n)/8}$. At the
critical radius, the gauge symmetry $SO(2n)\times U(1)$ is
enhanced to $E_{n+1}$. The electrically charged objects that are
enhancing the gauge symmetry have typically non-zero winding and
momentum, and quantum numbers under $SO(2n)$. The magnetically
charged objects with the same quantum numbers are identified as
BPS configurations with $SO(2n)$ monopoles on a Taub-NUT space. As
the radius $R$ is taken to be large they become heavier and
narrower as expected from their behavior as monopoles. We can
trace what these objects are from the Type I$'$ picture. The
spinor representation of $SO(2n)$ is given by a Taub-NUT space
where the order of the space corresponds to the monopole number
and the values of the blow up parameters serve as positions of
these monopoles. The adjoint representation of $SO(2n)$ is given
by fractional small Heterotic instantons. Such unusual Heterotic
background has a moduli space that is predicted to be the moduli
space of $E_{n+1}$ monopoles. Once again, the classical
singularity is smoothed out by the world-sheet instanton
corrections.
\subsection{Parameter mapping}

We conclude this Section by
 making the mapping of the Type I$^\prime$ configuration
discussed in Section 2 to the Heterotic set-up more explicit.
We will give a more precise computation of the masses for electrically
and magnetically charged objects.

The
Type I$^\prime$ background is given by \cite{witpol,BGL,theisen}
\bea e^{\phi}&\sim&(\Omega(x_9)/C)^5\cr
g_{\mu\nu}&=&\Omega^2(x_9)\eta_{\mu\nu} \eea{ee5} where
$\Omega(x_9)\sim C^{5/6}(B+(8-n)x_9))^{-1/6}$, and where, for
simplicity, all the $\pi$ and $\alpha^\prime$ factors have been
ignored. Here $x_9$ belongs to the interval $[0,\pi]$.

The
Type I$^\prime$ background is specified by the parameters $B$ and $C$.
The relation with the Heterotic parameters
$R,\phi_h$ is given by \bea &Re^{-2\phi_h}\sim D^5C^5\cr\nonumber
&DC^{5/3}\sim (8-n)^{1/2}[(B+2\pi (8-n))^{4/3}-B^{4/3}]^{-1/2}
\eea{ee6} where $D^2$ (a function of $B$ and $C$) is the factor that converts
 the 9 dimensional Heterotic metric to the Type I$^\prime$ metric.

The mass for a D0 brane stuck at the orientifold plane has been
worked
 out and compared with the Heterotic BPS formula in \cite{BGL,theisen}.
The result is \begin{equation} {M_{D0}\over 2} (=\sim
\Omega(0)e^{-\phi (0)})={1\over D}\left ({R\over\alpha^\prime}-
{(8-n)\over 8R}\right ) \eeq{ee7} The right hand side vanishes for
$R=R_{{\rm crit}}$: this tells us that the Type I$^\prime$
electric objects that are becoming massless at the enhanced
symmetry point are D0 branes stuck at the orientifold plane.
$R-R^2_{{\rm crit}}/R$ is identified with the $E_{n+1}$ Higgs VEV
in the Heterotic theory which enhances $SO(2n)\times U(1)$ to
$E_{n+1}$. We see that the mass for a $W^{\pm}$ bosons is given by
the Higgs VEV in agreement with the general expectations. The
factor of 1/2 takes into accounts the fact that the D0 brane is
stuck while $D$ takes into accounts the rescaling between the Type
I$^\prime$ and the Heterotic metrics.

We can now do a similar check for a stuck NS brane. On general
grounds, its mass should be given by the Higgs VEV divided by the
square of the $E_{n+1}$ coupling constant.  The tension for a
stuck NS brane is given by \begin{equation} T_{NS}\sim \int d^6
x{1\over g_s^2l_s^6}\sqrt{g_I}\sim \Omega(0)^6 e^{-2\phi (0)}
\eeq{ee8} Combining equations~(\ref{ee5}) with the first of
equations~(\ref{ee6}), we can compute the ratio \begin{equation}
{1\over g_{YM}^2}={T_{NS}\over M_{D0}}=\Omega(0)^5e^{-\phi
(0)}\sim C^5\sim {1\over D^5}Re^{-2\phi_h} \eeq{ee9} The factor
$D^5$ is just the effect of the metric rescaling. In Heterotic
units, we recover the result that the $E_{n+1}$ gauge coupling is
${1\over g_{YM}^2}=Re^{-2\phi_h}$, in agreement with the
perturbative analysis.

\section{Smoothness of the moduli space}
We expect the moduli space of monopoles to be a smooth manifold.
All the previous examples were focused on configurations where we
expected to get a smooth moduli space. As discussed in
\cite{Witten}, due to the equation \begin{equation}
\partial^2 \phi_h= \Tr F^2-\Tr R^2
\eeq{ed1} singularities in $\Tr F^2$ drive the Heterotic string
to a non-perturbative regime where we expect singularities in the
moduli space. Singularities in $\Tr R^2$, on the other hand,
 keep the theory in the perturbative
region and the only possible singularities may come from a break
down of the two dimensional sigma model description.
 It was argued in \cite{Witten} that in the
absence of gauge fields the sigma-model can not fail. In
\cite{Sen,Witten}, the criterion for having a finite coupling
constant was satisfied by considering no gauge fields at all. We
considered generalizations where the gauge fields are present but
still  $\Tr F^2$ is regular. The background parameters, such as
the radius of $S^1$ and the Wilson lines, were chosen in such a
way that the 9 dimensional gauge group $SO(2n)$ is spontaneously
broken to the Cartan sub-algebra. In this background, the only
objects (that give rise to 6 dimensional moduli) introduced in the
game were BPS monopoles of the space-time gauge group $SO(2n)$ and
KK monopoles. BPS monopoles have a regular $\Tr F^2$ while KK
monopoles may induce only a singularity in $\Tr R^2$.

It is fairly easy to consider singular configurations. Heterotic
background with non trivial gauge fields associated with
instantons suffer from small instanton singularities. Near the
singular points, the perturbative description breaks down, since,
due to equation~(\ref{ed1}), the dilaton is blowing up.
 The singularity survives quantum
corrections and it is associated with a restoration of a
six-dimensional gauge theory via Higgs mechanism. In the type
I$^\prime$ picture these configurations introduce extra D6 branes
wrapped along $S^1$. When D6 branes touch, some degrees of freedom
become massless and there is an enhanced gauge symmetry.

We can have more general non-perturbative vacua of the Heterotic
string with 6 dimensional tensor multiplets. Their existence can
be easily explained in the dual Type I theory, which can be
thought of as a Type II theory moded out by the world-sheet
parity. Background with tensor multiplets can be easily obtained
with a $Z_k$ orbifold. In Type II, each of the $k-1$ twisted
sectors give rise to a hypermultiplets
 and a tensor multiplet. The world-sheet parity may project out
the tensor multiplet or the hypermultiplet. The various consistent
possibilities are associated with the different types of gauge
bundles, depending whether they admit vector structure or not. In
such backgrounds, some of the blowing up modes of the ALE space
are projected out, and the space-time can not become completely
smooth. The configuration that we considered in this paper
corresponds to projecting out all the tensor multiplets and it
corresponds to the ``compactification'' on a smooth ALE space. The
various 6 dimensional models that may be obtained in Type I
orientifold constructions are discussed in
\cite{augustouno,augustodue,augustotre,gp,gj,ken,kentwo,kenthree}.
The type I$^\prime$ description was
considered in \cite{BK,hzsix,BK2}: the consistent models are
obtained by disposing, in a $Z_2$ symmetric way, $k$ NS branes on
the segment. A NS brane in the middle of the segment supports a
tensor multiplet, while a NS brane stuck at one of the orientifold
points supports a hypermultiplet. D6 branes can be stretched
between NS branes; they have the interpretation of small
fractional instantons in the Heterotic string. The number of D6
branes is fixed by RR space-time charge conservation; as shown in
\cite{hzsix}, this is completely equivalent to the anomaly
cancellation in the six-dimensional gauge theory. All these moduli
space have small instanton singularities at the origin of the
Higgs branch.

There is a natural mapping of all these six-dimensional configurations
to three-dimensional $N=4$ gauge theories, where mirror symmetry can be used
in order to extract information on the moduli space. For example, it was
explained in \cite{hzsix}, using the brane description, how the
moduli space of $n$ small $E_8$ instantons on an $A_k$ singularity
is mapped to the Coulomb branch of a three dimensional, $N=4$
supersymmetric, $U(k)$ gauge theory with $n$ flavors.

The configuration with all the NS branes stuck at one of the
orientifolds was considered in \cite{hzsix} only for a specific
example. We see that it generically represents the Type I$^\prime$
description of the Type I or Heterotic vacuum corresponding to a
``compactification'' on a smooth ALE space.

\section{$D_n$ singularities}

Let us briefly consider the case of $D_k$ singularities.

From the Heterotic point of view, we consider the moduli space of
$D_k$ ALF spaces. These spaces are asymptotic to (Taub-NUT)$/D_k$,
where $D_k$ is the binary dihedral group of order $4(k-2)$.

Unlike their cousins - the $A_k$ ALF spaces - that have the very
simple metric~(\ref{pot}) with the potential~(\ref{pottwo}), the
$D_k$ ALF space metric is much more complicated
\cite{kapuone,kaputwo,kaputhree,rocek}. Asymptotically, for large
$|x|$, their metric can be given in the form of
equation~(\ref{pot}) with a potential
\begin{equation}
V(\vec x)=1-{4\over |x|}+\sum_{i=1}^k{1\over |x-x_i|}
+\sum_{i=1}^k{1\over |x+x_i|}
\label{Dmetric}
\end{equation}
Some explicit construction for the full metric can be found in
\cite{kapuone,kaputwo,kaputhree}.

Consider, for simplicity, a Heterotic background without gauge
fields. When we tune the radius $R$ to the self-dual point, we
still expect to find an $SU(2)$ symmetry in the spectrum. With
respect to the Taub-NUT space, the $D_k$ ALF spaces have an
additional $Z_2$ action that combines with the cyclic group $C_k$
to produce the dihedral group $D_k$. This $Z_2$ is manifest in
equation~(\ref{Dmetric}) and acts non-trivially on the space
$R^3$. As a result, the nine-dimensional $SU(2)$ gauge theory is
moded out by a $Z_2$ that changes sign to three of the nine
space-time coordinates. The nine-dimensional Lorentz invariance is
explicitly broken; we are dealing with an orbifold gauge theory,
or, in other words, with a theory with impurities. The moduli
space of $D_k$ ALF spaces is then identified with the moduli space
of $k$ monopoles of the orbifolded $SU(2)$ gauge theory. We can
easily realize the same monopole configuration in terms of branes;
for example, with D3 branes stretched between two NS branes in the
presence of an orientifold plane O3$^-$. The corresponding
monopole moduli space is then identified with the Coulomb branch
of an N=4 three-dimensional $SO(2k)$ gauge theory by using the
standard rules from \cite{HW}. The relation between the moduli
space for Heterotic on $D_k$ ALE singularities and the Coulomb
branch of this gauge theory has been conjectured in \cite{Witten}.

The Type I$^\prime$ description is simple. As discussed in
\cite{HZissues}, a $D_k$ singularity is modeled in the dual
picture using NS branes in the presence of an $ON^0$ plane. There
are two $ON^0$ planes, each at the endpoints of the Type I$'$
interval.  The presence of an O8$^-$ plane induces in addition an
O6 plane. There are six-dimensional fields living on the $ON^0$
planes; if we choose an O6$^-$ plane, there is a hypermultiplet on
the $ON^0$ plane \cite{HZissues}.

\begin{figure}[h]
  \centering
  \resizebox{5cm}{!}{\includegraphics{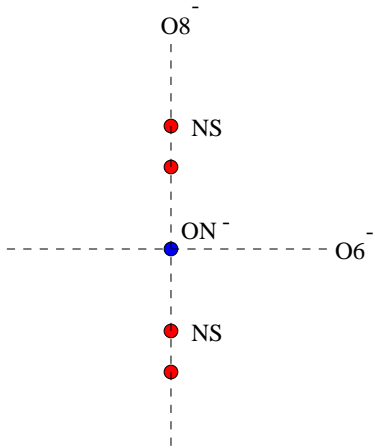}}
  \caption{$D_k$ monopoles is Type I$'$. The vertical dashed line
  represents an $O8^-$ plane. The horizontal dashed line represents
  an $O6^-$ plane. At the intersection of both there is an $ON^-$ plane.
  $2k$ half NS branes are placed around the $ON$ plane in a symmetric fashion.
  As for the $A_k$ case half NS branes are confined to the $O8^-$
  plane.
  }\label{On}
\end{figure}

We will consider only one of the $ON^0$ planes, as the physics is
confined to one of the $O8^-$ planes. It is more convenient, for
our purposes, to think of an $ON^0$ plane as the combination of a
$ON^-$ plane with a physical NS brane. We can put extra $2k-2$
stuck NS branes on the O8$^-$, as in figure \ref{On}. We think of
this system as an $ON^-$ plane with $2k$ half NS branes stuck to
the $O8^-$ plane \cite{HZissues}. Since there is an O6$^-$ plane,
each of the half NS branes has an image on the O8$^-$ plane. We
have a total of $k$ hypermultiplets that parameterize our moduli
space. There is a cosmological constant in the bulk that prevents
the NS branes from leaving the orientifold plane. As a
consequence, there is no singularity associated to their motion in
the bulk and the moduli space is smooth.

There is a potential weak point in this argument due to the fact
that two half NS branes in the game arise in a perturbative
description of the $ON^0$ plane as twisted states. The difficulty
stems from the fact that we do not have a satisfying description
for the $ON^-$ plane whereas the $ON^0$ plane admits a
perturbative description. The two half NS branes related to the
$ON^0$ plane however do not leave the $O8^-$ plane by the same
argument of charge conservation valid for the other NS branes. As
an example, we can consider the case $k=2$. The Heterotic dual
contains a $D_2$ singularity. Since $D_2$ is the product of two
disjoint $A_1$ singularities, we expect that the moduli space is
the product of two Atiyah-Hitchin manifolds. For $k=1$ the group
$D_1$ is $SO(2)$ and the configuration is just an $ON^0 (= ON^- +$
NS) plane located at the intersection of the $O8^-$ and $O6^-$
planes. The moduli space is flat space as expected from half a NS
brane and its image as they leave the $ON^-$ plane.

In this way, the Type I$^\prime$ picture
suggests that the moduli space of orbifolded gauge theories is
smooth despite the singularity in the space where the monopoles
are living. We can also formulate a related conjecture for
three-dimensional gauge theories: the Coulomb branch of N=4 SYM
$SO(2k)$ theories is a smooth manifold.

The generalization to include D8 and D6 branes is straightforward.

\section{Conclusions}

We provided an explicit construction of $E_{n+1}$ monopoles in
string theory. Our investigation leads us to identify the
monopoles as Half NS branes stuck at an O8 plane in Type
I$^\prime$. These NS branes transform in the spinor representation
of an $SO(2n)$ subgroup of $E_{n+1}$. Conservation of RR charge
does not allow the NS branes to move outside the orientifold
plane, giving a stringy interpretation for the smoothness of the
monopole moduli space. The same argument can be used to predict
smoothness in the moduli space of more general and less studied
gauge theory monopoles.

We also connected our configuration with Heterotic backgrounds
where monopoles naturally appear in the form of KK monopoles. We
were able to give an explanation and to generalize some results on
the Heterotic moduli space appeared in the literature
\cite{Sen,Witten}. We did not discuss the appearance of
three-dimensional gauge theories in this context as they are very
natural in view of our discussion and the results of \cite{CH,HW}.

In this paper, we just considered an example of a
 particular class of monopoles,
which, nevertheless, has many ramifications and connections with
different string backgrounds. Many other monopole configurations
should naturally show up in string theory. We already know of
monopole moduli spaces appearing in gauge and string theories in
many contexts, stemming from three-dimensional gauge theories to
singular Heterotic backgrounds. Every time an Atiyah-Hitchin
manifold shows up in the string moduli space, it is worthwhile to
look around for monopoles.

A natural direct extension of our investigation
would be to study systems with D7 branes. Exceptional groups
also appear quite naturally in this context. A chain of dualities
would lead us to consider more general Heterotic backgrounds
and F-theory models.

\vskip .1in {\bf Acknowledgements}\vskip .1in \noindent We would
like to thank Nissan Yitzhaki for useful discussions and comments.
A. Z. would like to thank the Institute of theoretical Physics,
Santa Barbara, where this work was initiated, for its kind
hospitality. A. H. and A. Z. are partially supported by the
National Science Foundation under grant no. PHY94-07194. A. H. is
supported in part by the DOE under grant no. DE-FC02-94ER40818, by
an A. P. Sloan Foundation Fellowship and by a DOE OJI award. A. Z.
is partially supported by INFN and MURST, and by the European
Commission TMR program ERBFMRX-CT96-0045, wherein A. Z. is
associated to the University of Torino.

\bibliographystyle{utphys}
\bibliography{monopoles}
\end{document}